\documentclass[conference]{IEEEtran}
\usepackage{cite}
\usepackage{booktabs}
\usepackage{float}
\usepackage{mathtools}
\usepackage{algorithm2e}
\usepackage{xcolor}
\usepackage{listings}
\usepackage[T1]{fontenc}
\usepackage{caption}
\usepackage{graphicx}
\usepackage{tabularx}
\restylefloat{table}
\begin{document}

	\title{Performance Evaluation of Deep Learning Tools in Docker Containers}
	
	\author{\IEEEauthorblockN{Pengfei Xu}
		\IEEEauthorblockA{Department of Computer Science\\	
			Hong Kong Baptist University\\
			Email: pengfeixu@comp.hkbu.edu.hk}
		\and
		\IEEEauthorblockN{Shaohuai Shi}
		\IEEEauthorblockA{Department of Computer Science\\
			Hong Kong Baptist University\\
			Email: csshshi@comp.hkbu.edu.hk}
		\and
		\IEEEauthorblockN{Xiaowen Chu}
		\IEEEauthorblockA{Department of Computer Science\\
			Hong Kong Baptist University\\
			Email: chxw@comp.hkbu.edu.hk}}
	\maketitle
	
	\begin{abstract}
	With the success of deep learning techniques in a broad range of application domains, many deep learning software frameworks have been developed and are being updated frequently to adapt to new hardware features and software libraries, which bring a big challenge for end users and system administrators. To address this problem, container techniques are widely used to simplify the deployment and management of deep learning software. However, it remains unknown whether container techniques bring any performance penalty to deep learning applications. The purpose of this work is to systematically evaluate the impact of docker container on the performance of deep learning applications. We first benchmark the performance of system components (IO, CPU and GPU) in a docker container and the host system and compare the results to see if there's any difference. According to our results, we find that computational intensive jobs, either running on CPU or GPU, have small overhead indicating docker containers can be applied to deep learning programs. Then we evaluate the performance of some popular deep learning tools deployed in a docker container and the host system. It turns out that the docker container will not cause noticeable drawbacks while running those deep learning tools. So encapsulating deep learning tool in a container is a feasible solution.
	\end{abstract}

	\IEEEpeerreviewmaketitle
	
	\section{Introduction}
	Ever since the great success of deep learning techniques in many application domains, more and more deep learning software tools have been developed by different research institutions and companies for both academic research and commercial use \cite{shi2016benchmarking}. Popular tools like Caffe\cite{jia2014caffe}, CNTK\cite{yu2014introduction}, MXNet\cite{chen2015mxnet}, TensorFlow\cite{abadi2016tensorflow}, Torch\cite{collobert2011torch7}, etc. are still being actively developed and their new versions are being released frequently, which brings significant software management challenge to system administrators. It is even worse when different tools or different versions of the same tool need to be installed in a system that is shared by multiple users. A practical solution to simplify the management of deep learning tools is to make use of docker containers so that environmental setting conflicts can be easily resolved by packaging a software and its all required libraries into a single image \cite{fu2016toward}. Despite its popularity in practical usage, there lacks a systematic analysis on the performance overhead brought by docker containers for deep learning tools. This paper aims to investigate the impact of docker containers on the performance of deep learning tools.

	A typical deep learning training workflow involves data access from/to disk drives and intensive data processing on CPUs and/or accelerators such as GPUs \cite{chetlur2014cudnn}. Therefore we evaluate the performance of CPU, GPU, disk I/O, and the overall deep learning training with and without dock container, respectively. For CPU performance, we make use of two classical and representative benchmarks, HPL and HPCG. For GPU performance, a set of GPU programs are selected to test different types of GPU operations. Disk I/O performance can be another important factor when huge amount of data are fed to neural networks during training process \cite{lecun2015deep}. We test I/O performance from several aspects, including I/O access latency, random access throughput, and sequential access throughput. At last, we evaluate the training performance of five popular deep learning software tools with different neural network models and datasets. Based on our experimental results, we find that docker containers have negligible overhead in computing-intensive tasks on both CPU and GPU. The I/O performance of sequential access under the docker container is found to be at the same level as the host system. When it comes to random access, we observe even shorter response time on docker container than on the host system using one of the tested disk drives. This is because docker containers can make better use of the NAND cache on the hard disk to gain faster random data access. Since each factor mentioned above has satisfactory results on docker, it is not surprising to find that running deep learning tools in docker containers has negligible overhead compared to running on host systems directly. 
	
	This paper is organized as follows. Background of deep learning and docker containers and related work are introduced in Section \ref{background}. The design of our experiments is presented in Section \ref{experiment}. We show our experimental results and analysis in Section \ref{results}. We conclude our work in Section \ref{conclusion}.

	\section{Background}\label{background}
	\subsection{Deep Learning}
	Deep learning is a class of machine learning techniques which powers great number of facets in our everyday life. Deep neural networks are built of many processing layers and are able to learn features from a mass of data with various stages of abstraction \cite{lecun2015deep}. This technology has many applications like speech recognition\cite{hinton2012deep}\cite{dahl2012context}\cite{deng2013new}\cite{graves2013speech}, image recognition\cite{he2016deep}\cite{nguyen2015deep}\cite{krizhevsky2012imagenet}\cite{deng2009imagenet}, natural language processing\cite{collobert2008unified}\cite{manning2014stanford}\cite{collobert2011natural}, and the list is getting longer and longer. Comparing with conventional machine leaning techniques, deep learning has less limitation on the data fed to the computer to learn\cite{lecun2015deep}. But training a deep neural network for a certain problem is not an easy task and it requires significant computational power.
	
	To this end, many-core parallel processors like GPUs are widely used to facilitate deep learning tasks \cite{chetlur2014cudnn}. Some stages of deep learning process can be eventually mapped to linear algebra operations which can usually be efficiently implemented on parallel processors. As a matter of fact, many popular deep learning software such as Caffe\cite{jia2014caffe}, CNTK\cite{yu2014introduction}, MXNet\cite{chen2015mxnet}, TensorFlow\cite{abadi2016tensorflow}, and Torch\cite{collobert2011torch7}, have all implemented the support of GPUs whose performance is significantly better than CPUs\cite{shi2016benchmarking}.
	\subsection{Docker Container}
\begin{figure}[!ht]
		\centering
		\includegraphics[width=0.32\textwidth]{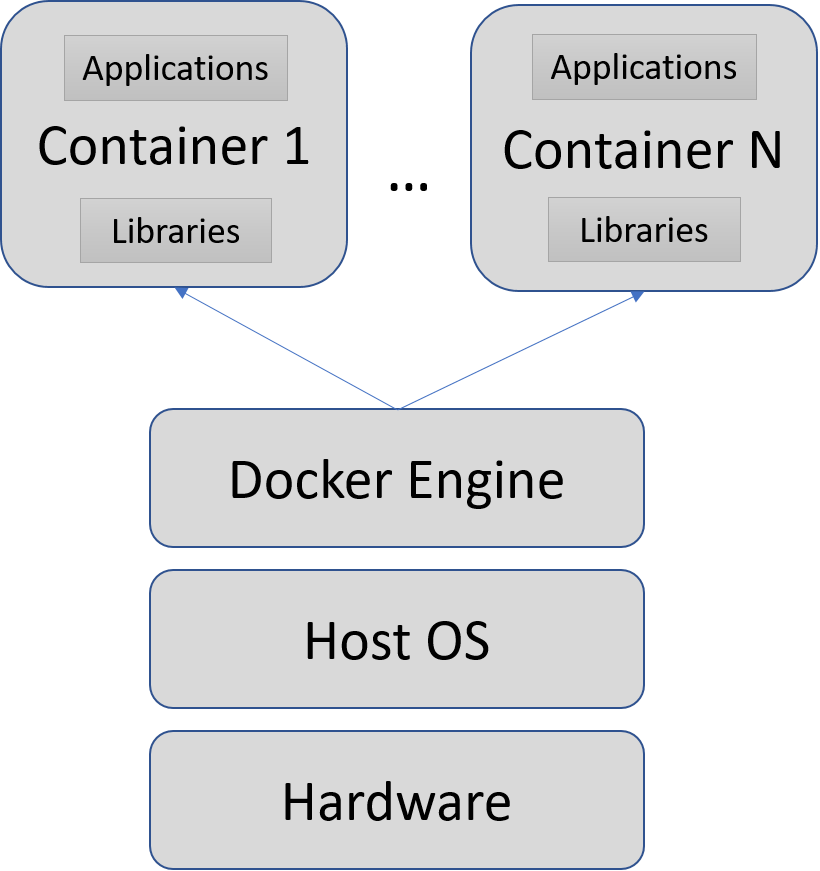}
		\caption{Docker Architecture}
		\label{fig:docker}
	\end{figure}

	Docker is a container virtualization technology which behaves similar to a light-weighted virtual machine, and it is the most popular open source application-oriented approach\cite{plauth2017performance}. Docker isolates each independent container running on the same instance of operating system by making use of Linux kernel features like control groups and namespaces\cite{plauth2017performance}. A simple illustration of docker can be found in Figure \ref{fig:docker}.
	The simple architecture of docker leads to less overhead as compared to other multi-layered types of virtualization technology. Each docker container encapsulates an application and its required dependencies, and can be run on different machines on top of a docker engine. Docker images can also be easily shared and distributed once they have been built.
	\subsection{Related work}
	Docker as a light-weighted virtualization solution has drawn attentions from the research community to explore its potential in different applications. A few studies have done great jobs in comparing the performance of docker with other types of virtualization solution \cite{kamarainen2015virtual}\cite{beserra2016performance}\cite{chung2016using}. 
	
	It has been shown that the docker container outperforms QEMU\cite{kamarainen2015virtual} when running popular GPU gaming benchmark, and it runs as good as native OS. The purpose of \cite{kamarainen2015virtual} is to study the possibilities of hosting cloud gaming service using docker containers. We can only see the overall gaming performance from it. Another work \cite{haydel2015enhancing} also tests the performance of GPUs in docker containers and VMs by running the algorithm SGEMM which performs matrix multiplication followed by an addition operation: $C = \alpha A \times B + \beta C$. It indicates that invoking GPUs in docker containers will not cause much overhead. Others also put docker into high performance computing conditions\cite{chung2016using} and compare it with virtual machines. Chung's work\cite{chung2016using} evaluates the performance by running HPL benchmark which is compiled with OpenMPI and OpenBLAS and run with different problem sizes. According to their experimental results, docker container manages to have better performance than VMs while using less RAM. Docker also has better scalability than VMs. As the number of VM instances and docker containers increases, docker can keep its performance without dramatically losing computing ability due to large overhead like VMs do. Different from the previous work, we aim to quantitatively measure the performance overhead caused by docker containers when running deep learning software.

	\section{Experimental Design}\label{experiment}
	In this section, we will introduce the design of experiments to compare the performance between using docker container and without docker container. Four groups of experiments (i.e., CPU, GPU, disk I/O, and deep learning tools) are designed to evaluate the performance differences of the same task running in docker container environment and in host system directly. To make a fair comparison, we make sure that the irrelevant variables like compiler versions and software libraries are the same in docker container and host system. All experiments are performed on two hardware platforms: a desktop PC and a rack server. Detailed information of our hardware configurations can be found in Table \ref{gpu-config} and \ref{server-config}. All reported results are the average of 20 runs unless otherwise specified.
	\begin{table}[!ht]
		\centering
		\caption{Hardware Configuration of Desktop PC Platform}
		\label{gpu-config}
		\begin{tabular}{@{}ll@{}}
			\toprule
			Item        & Model                		\\ \midrule
			CPU         & Intel i7-6800K       		\\
			Motherboard & ASUS X99-A II        		\\
			RAM         & Kingston 64G DDR4         \\
			GPU1        & Nvidia GTX TITAN X 		\\
			GPU2        & Nvidia GTX 980 			\\
			Hard Drive	& Seagate ST3000DM008 7200RPM HDD    \\
			\bottomrule
		\end{tabular}
	\end{table}

	\begin{table}[!ht]
		\centering
		\caption{Hardware Configuration of Server Platform (Lenovo x3650 M5)}
		\label{server-config}
		\resizebox{\textwidth/2}{!}{
		\begin{tabular}{@{}ll@{}}
			\toprule
			Item        & Model                \\ \midrule
			CPU         & Intel Xeon E5-2620 v3      \\
			RAM         & Samsung 32G DDR4             \\
			Hard Drive	& Lenovo System X 10K RPM 600GB (SAS 12Gb)    \\
			\bottomrule
		\end{tabular}
	}
	\end{table}
	Since we will perform our experiments with NVIDIA GPUs, we use NVIDIA docker\footnote{Details about NVIDIA docker: https://github.com/NVIDIA/nvidia-docker} which is a thin wrapper on top of docker. When we start the NVIDIA docker, it calls the docker and relies on NVIDIA Docker plugin to load GPU driver and communicate with GPUs directly. NVIDIA docker only changes the behavior of \textit{docker run} and \textit{docker create} commands as stated in its official document.
	\subsection{CPU Performance}
	In order to test the performance of CPU, we run both HPL and HPCG benchmarks from Intel MKL library in \textit{Intel Parallel Studio 2017 Update 2}. Both of them are measured in GFlops (Giga floating point operations per second).
	
	For HPL benchmark, the problem sizes are set from 2,000 to 45,000, simulating different levels of computational intensity. The detailed experimental settings are shown in Table \ref{hpl-config}.
	\begin{table}[!ht]
		\centering
		\caption{HPL Configuration}
		\label{hpl-config}
		\begin{tabular}{@{}cc@{}}
			\toprule
			Problem Size  & Leading Dimension   \\ \midrule
			2000  & 2000  \\
			5000  & 5008  \\
			10000 & 10000 \\
			15000 & 15000 \\
			18000 & 18008 \\
			20000 & 20016 \\
			22000 & 22008 \\
			25000 & 25000 \\
			26000 & 26000 \\
			27000 & 27000 \\
			30000 & 30000 \\
			35000 & 35000 \\
			40000 & 40000 \\
			45000 & 45000 \\
			\bottomrule
		\end{tabular}
	\end{table}
	HPL is originated from Linpack benchmark that measures the floating-point performance by solving a linear system of equations of order \textit{n} (i.e., problem size) \cite{dongarra2004introduction}:
	\begin{equation*}\label{lin}
	Ax = b;  A \in \Re^{n\times n};  x,b \in \Re^n
	\end{equation*}
	To solve this linear system, it first computes the LU factorization with row partial pivoting of the $n$-by-$n+1$ coefficient matrix:
	\begin{equation*}
	[A\ b] = [[L,U] y]
	\end{equation*}
	Since the lower triangular factor $L$ is applied to $b$ as the factorization progresses, the solution $x$ is obtained by solving the upper triangular system $Ux = y$. The lower triangular matrix $L$ is left unpivoted and the array of pivots is not returned\cite{dongarra2004introduction}. HPL serves as system stress test due to its intense computing property.
	
	HPCG is another popular benchmark designed for HPC systems to be closer to real application. Basically, high performance conjugate gradient (HPCG) is consisted of computations and data access patterns which are more commonly seen in real applications\cite{dongarra2013toward}. This benchmark program also solves a linear system \begin{math}
	Ax = b
	\end{math}, but with a conjugate gradient method. As mentioned in \cite{dongarra2015hpcg}, a system that is designed for good HPL performance can result in wrong choices for adding unnecessary components or complexity to the system. We run HPCG with the problem dimension of 192 and running time being 1800 seconds to get valid benchmark results.

	\subsection{GPU Performance}
	As for comparing GPU performance, we choose some representative applications from CUDA 8.0 samples. Each of them performs different operations on GPU.
	
	First we test the effective data transmission throughput from CPU to GPU, GPU to CPU, and GPU to GPU, aiming to check if docker container will affect the speed of data transmission which is crucial in training neural networks as huge amount of training data need to be delivered between CPU and GPU. 
	
	Then we perform a convolution operation on a $18432 \times 18432$ image, which is commonly used in deep learning to extract features from training data like images\cite{sze2017efficient}, and measure the throughput in mega-pixels per second.
	
	Matrix multiplication is our third GPU application, as it is widely used in not only fully-connected layers, but also convolutional layers \cite{sze2017efficient}. So the efficiency of performing matrix multiplication on GPU is crucial to deep learning performance. We generate matrix $A$ of size $23040 \times 17280$ and matrix $B$ of size $17280 \times 11520$ and calculate matrix $C$ by calling CUBLAS function:
	\begin{equation*}
	A \times B = C
	\end{equation*}
where
	\begin{equation*}
	A \in \Re^{23040 \times 17280}, B \in \Re^{17280 \times 11520}, C \in \Re^{23040 \times 11520}.
	\end{equation*}
	Such sizes are selected to fill the 4GB memory of GTX 980 so as to make the workload sufficiently large.
	
	Next, we benchmark the performance of self-defined kernels by calculating 64-bin and 256-bin histogram of 67108864 random numbers ranging from 0 to 255. The pseudo-code of the algorithm is shown in Algorithm 1. The CUDA kernel function divides the data into many individual parts. Each thread processes one part and stores the sub-histogram in its own storage space. Then it merges all sub-histograms to get the final result \cite{podlozhnyuk2007histogram}.

	\begin{algorithm}[ht]\label{histoalgo}
		\KwIn{Random number array \textbf{data}}
		\KwOut{Histogram array \textbf{result}}
		\For{counter \textless\ BIN\_COUNT}
		{
			result[counter] = 0;
		}
		\For{counter \textless\ number of data}
		{
			result[data[counter]]++;
		}
		\caption{Histogram Calculation}
	\end{algorithm}

	\begin{figure*}[h]
	\centering
	\includegraphics[width=0.9\textwidth]{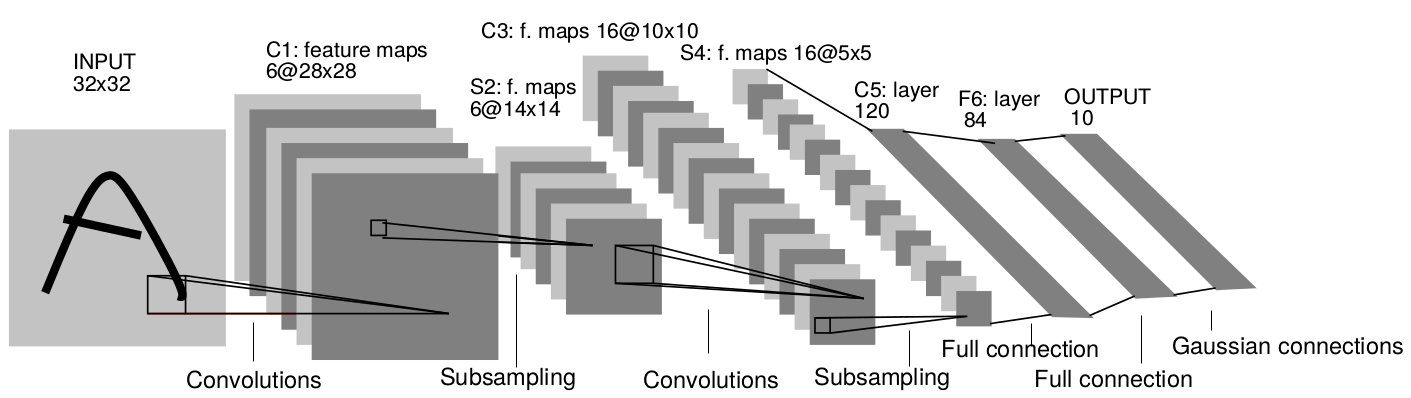}
	\caption{A Simple Example of CNN(LeNet)\cite{lecun1995comparison}}
	\label{lenet}
	\end{figure*}

	Lastly, we test different versions of matrix transpose which are memory intensive \cite{mei2014benchmarking}\cite{mei2017tpds}. We consider 8 different approaches including simple copy, simple copy with shared memory, naive transpose, coalesced transpose, shared memory bank conflicts, decomposing transpose, partition camping, and diagonal block reordering from \cite{ruetsch2009optimizing}. We use matrix transpose to test the efficiency of GPU memory access \cite{ruetsch2009optimizing}. Matrix transpose operations are also commonly used in deep learning tools \cite{shi2017transpose}.
	
	\subsection{I/O Performance}
	In this part we test disk I/O performance in different ways with \textit{dd} and \textit{ioping} tools. \textit{dd} is used to write multiple files with different sizes to hard disk and we can get access speed from the outputs of \textit{dd} directly. Below shows an example of writing out a 1 GB file:
	\begin{lstlisting}[language=bash]
dd if=/dev/zero of=bigfile bs=64k \
	count=16k conv=fdatasync
	\end{lstlisting}
	Basically, this command reads a stream of null characters in blocks and each block contains 64KB of data. Then it physically writes to an output file named \textit{bigfile} to make I/O operations. There are 16,000 blocks, which mean that in total
	\begin{math}
	16,000 \times 64KB = 1GB
	\end{math} of data are written to the hard drive for measuring the sequential writing throughput. We can use the number of blocks, i.e., the \textit{count} flag, to control the file size. Specifically, we set \textit{count} to 16, 1600, 8k, and 16k to test with files of size 1MB, 100MB, 512MB, and 1GB respectively.
	
	Then we use \textit{ioping} to benchmark different access types including cache I/O random access test, direct I/O random access test, as well as I/O latency test. Details are listed in Table \ref{iopingtest}.
	\begin{table}[!ht]
	\centering
	\caption{\textit{ioping} Tests}
	\label{iopingtest}
	\begin{tabular}{ll}
		\toprule
		Purpose           			& Bash cmd    \\ \midrule
		IO latency                  & ioping -c 10 .       \\
		Cached IO random access     & ioping -C -R -w 5 .  \\
		Direct IO random access     & ioping -D -R -w 5 .  \\
		\bottomrule
	\end{tabular}
	\end{table}

	\subsection{Performance of Deep Learning Tools}
	After measuring the performance of individual factors, we come to evaluate the training performance of three representative neural networks: Fully Connected Networks (FCNs), Convolutional Neural Networks (CNNs), and Recurrent Neural Networks (RNNs) using different deep learning tools, including Caffe\cite{jia2014caffe}, CNTK\cite{yu2014introduction}, MXNet\cite{chen2015mxnet}, TensorFlow\cite{abadi2016tensorflow}, and Torch\cite{collobert2011torch7}. The tested software versions are shown in Table \ref{dltool}. We measure the speed in unit of second per batch.
	\begin{table}[H]
		\centering
		\caption{Deep Learning Tools}
		\label{dltool}
		\begin{tabular}{@{}llll@{}}
			\toprule
			\multicolumn{1}{l}{Software} & Major Version & GitHub Commit ID & cuDNN \\ \midrule
			Caffe                        & 1.0           & 39f28e4          & v5.1  \\
			CNTK                         & 2.0           & 1ae666d          & v5.1  \\
			MXNet                        & 0.9           & 32dc3a2          & v5.1  \\
			TensorFlow                   & 1.0           & 4ac9c09          & v5.1  \\
			Torch                        & 7             & 748f5e3          & v5.1  \\ \bottomrule
		\end{tabular}
	\end{table}

	\subsubsection{FCN}
\begin{figure}[H]
		\centering
		\includegraphics[width=0.2\textwidth]{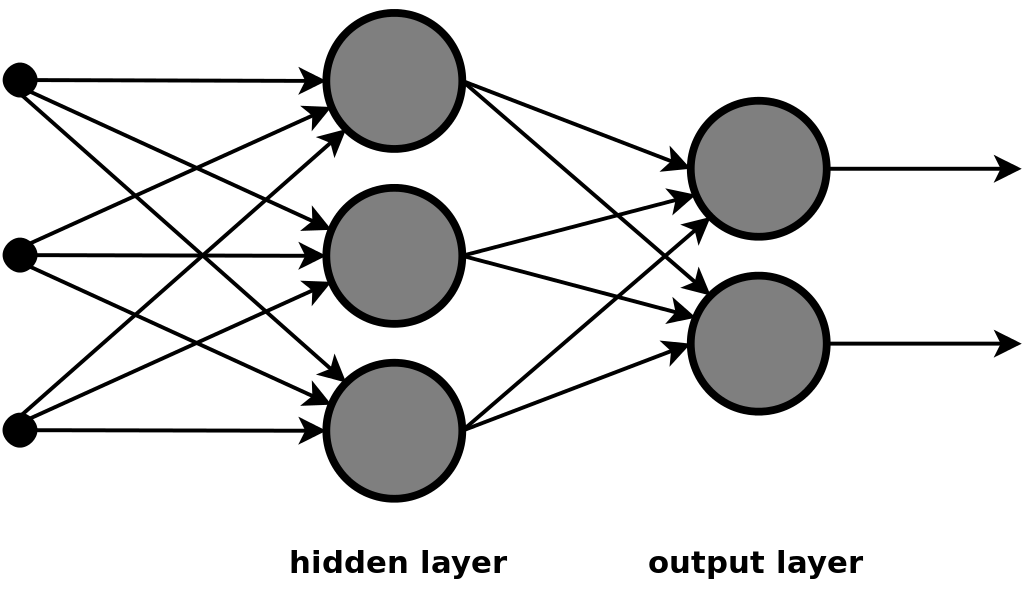}
		\caption{Fully Connected Network}
		\label{fcnpic}
	\end{figure}

	Fully connected networks are the simplest neural network model. Each neuron performs a simple forward active function and sends the result to all the neurons in the next layer (see Fig. \ref{fcnpic}).

	We design a FCN with 5 layers (including the input and output layers) whose configuration is shown in Table \ref{fcn5config}. We name it FCN5 and train it with MNIST dataset\cite{lecun1998mnist} which consists of 60000 labeled handwriting images.
	\begin{table}[!ht]
		\centering
		\caption{FCN5 Configuration}
		\label{fcn5config}
		\begin{tabular}{@{}lll@{}}
			\toprule
			Layer    & \# of nerons & Active function \\ \midrule
			Input    & 784           & -               \\
			Hidden 1 & 2048          & Sigmoid         \\
			Hidden 2 & 4096          & Sigmoid         \\
			Hidden 3 & 1024          & Sigmoid         \\
			Output   & 10            & Softmax         \\ \bottomrule
		\end{tabular}
	\end{table}
	\subsubsection{CNN}
	Convolutional Neural Networks (CNNs) are a set of models inspired by biology studies to simulate the way animal brains process images by introducing convolutional layers in artificial neural networks. As shown in Figure \ref{lenet}, in front of fully connected layers several convolutional layers are used to extract features from input images so that the classifier can better distinguish them with different labels.   	
	In this part, we select AlexNet\cite{krizhevsky2012imagenet} and ResNet\cite{he2016deep} to train Cifar10 dataset\cite{cifardatasets}.
	\subsubsection{RNN}
	Recurrent Neural Networks (RNNs) are widely used in applications like speech recognition, machine translation, language modeling, etc.\cite{zaremba2014recurrent}. Long short-term memory (LSTM) network is one of the most commonly used types in this category.
	\begin{figure}[H]
		\centering
		\includegraphics[width=0.45\textwidth]{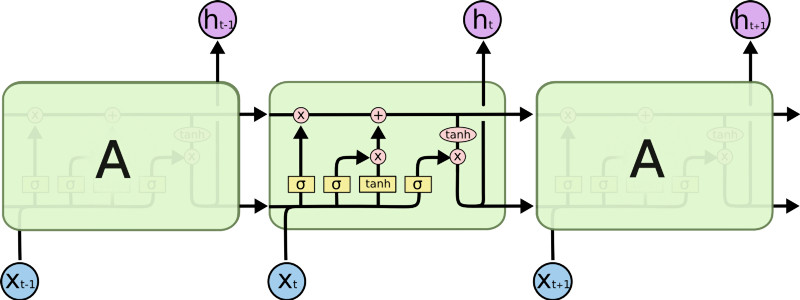}
		\caption{Example of LSTM layer}
		\label{lstm}
	\end{figure}
	As Figure \ref{lstm} illustrates, each LSTM layer packs a few sub-models and forms very deep neural networks by putting them together. In our experiments, we build a 2-layer LSTM network and train with PTB dataset. Each layer has 256 states and input texts are pre-processed into a character sequence with the length of 32.
	
	In summary, there are three categories of neural networks and four models will be included in our experiments. We will also train each model with different batch sizes to simulate the real training environment. Our experimental configurations are shown in Table \ref{deeplearingtoolsexperimentdesign}.
	
	\begin{table}[H]
		\centering
		\caption{Deep Learning Tools Experiment Design}
		\begin{tabular}{r|lcc}
			\toprule
			\multicolumn{1}{l}{Network Type} & Model & \multicolumn{1}{l}{Dataset} & \multicolumn{1}{l}{Batch Size} \\
			\midrule
			\midrule
			& FCN5  & MNIST & 512 \\
			\multicolumn{1}{l|}{FCN} & FCN5  & MNIST & 1024 \\
			& FCN5  & MNIST & 2048 \\
			\midrule
			& Alexnet & Cifar10 & 256 \\
			& Alexnet & Cifar10 & 512 \\
			\multicolumn{1}{l|}{CNN} & Alexnet & Cifar10 & 1024 \\
			\cmidrule{2-4}          & Resnet50 & Cifar10 & 16 \\
			& Resnet50 & Cifar10 & 32 \\
			& Resnet50 & Cifar10 & 64 \\
			\midrule
			& LSTM  & PTB   & 128 \\
			\multicolumn{1}{l|}{RNN} & LSTM  & PTB   & 256 \\
			& LSTM  & PTB   & 512 \\
			\bottomrule
		\end{tabular}%
		\label{deeplearingtoolsexperimentdesign}%
	\end{table}
	
	\section{Experimental Results and Analysis}\label{results}
	In this section, we will present our detailed experimental results. We mainly focus on presenting the performance difference between using docker container and without using docker container. We introduce the symbol $Diff_\%$ to represent the difference which is defined as follows:
	\begin{equation}
	Diff_\% = \frac{H-D}{H} \times 100\%
	\end{equation}
	where $H$ is the result without using docker container and $D$ is result of using docker container. 	\subsection{CPU Performance}
	\subsubsection{HPL Benchmark}
\begin{figure}[!ht]
		\centering
		\includegraphics[width=0.55\textwidth]{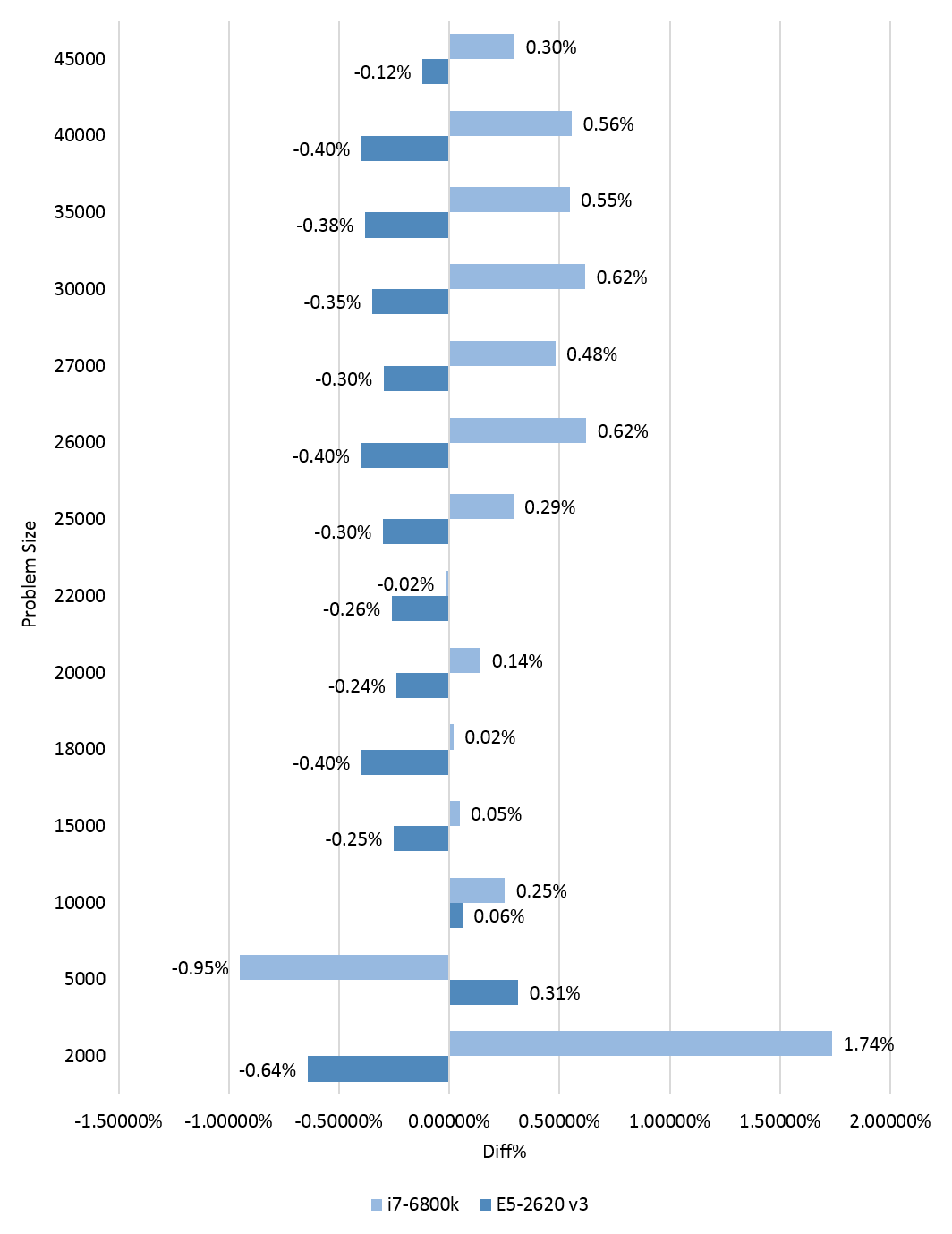}
		\caption{HPL Benchmark Results}
		\label{hplresult}
	\end{figure}

	We run HPL experiments on both Intel Xeon E5-2620 v3 and Intel i7-6800K platforms with different workloads. We gradually increase the problem size from 2,000 to 45,000. As illustrated in Fig. \ref{hplresult}, HPL tests on CPU performs nearly the same in docker container and host system in general. The differences are mostly less than 1\%. It is very interesting to notice that the HPL performance of using docker container can be even better on our server platform.
	
	\subsubsection{HPCG Benchmark}
	HPCG benchmark experiments also show that there is little overhead when using docker container. Table \ref{hpcgresult} illustrates that the differences between docker container and host system are merely 0.24425\% with E5-2620 v3 and 0.55624\% with i7-6800k respectively.
	\begin{table}[H]
		\centering
		\caption{HPCG Benchmark Results}
		\label{hpcgresult}
		\begin{tabular}{@{}lllll@{}}
			\toprule
			& Docker  & Host    & Difference & \begin{math}
			Diff_\%
			\end{math} \\ \midrule
			E5-2620 v3 & 5.525 & 5.539 & 0.013    & 0.244\%        \\
			i7-6800k   & 5.288 & 5.317 & 0.029    & 0.556\%        \\ \bottomrule
		\end{tabular}
	\end{table}
	Above all, minor performance differences are found in both HPL and HPCG benchmark experiments meaning that using CPU in a docker container won't introduce much overhead.
	\subsection{GPU Performance}
\begin{figure}[!ht]
		\includegraphics[width=0.5\textwidth]{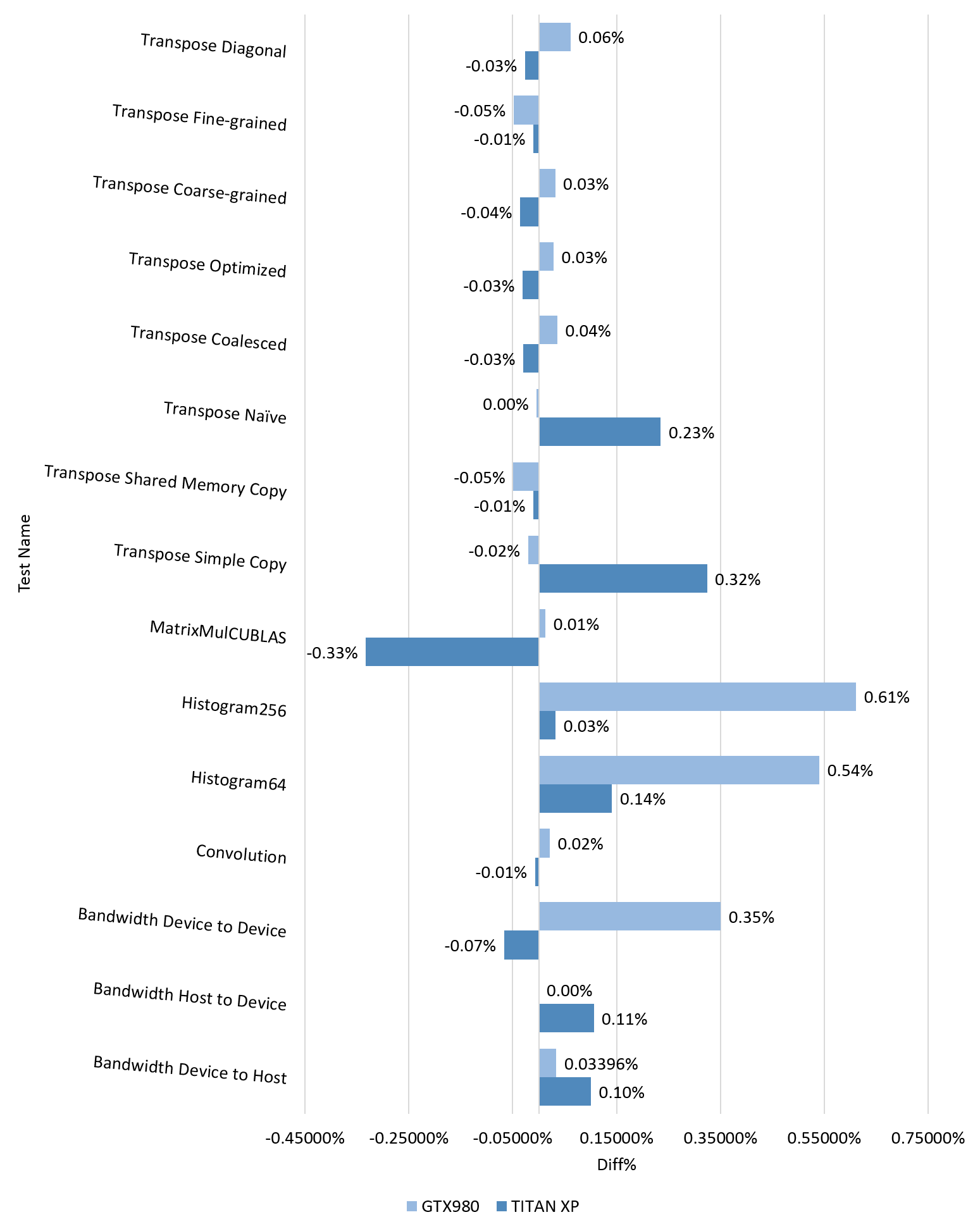}
		\caption{GPU Benchmark Results}
		\label{gpuresult}
	\end{figure}

	We perform our GPU tests on two generations of NVIDIA GPUs which are built in different architectures: the latest GTX Titan X (Pascal) and GTX 980 (Maxwell). GPU performance is a crucial factor for all the deep learning tools shown in Table \ref{dltool}. The experimental results are summarized in Fig. \ref{gpuresult}.
	
	We can see that the bandwidth tests show tiny performance difference in all three data transmission experiments. The most widely used matrix operation, matrix multiplication, also shows little performance difference. As for self-defined kernels like the histogram experiments, it is shown that the host system does perform better than in the docker container, especially when running on GTX980. A set of matrix transpose operations serves as a comprehensive evaluation of utilizing GPU for general purpose computation. Again there is no huge performance difference found in this set of experiments. GPU performance varies because of many physical environmental factors such as temperature change. Overall, it is quite promising that the maximum absolute $Diff_\%$ value is only 0.61\% (in the experiment of histogram256). These most frequently used operations in deep learning processes like data transmission, matrix multiplication, convolution operation, and matrix transpose all perform very well under docker container.
	\begin{table*}[h]
		\centering
		\caption{Deep Learning Tools Benchmark Batch Time $Diff_\%$ }
		\label{dlresult}
		\resizebox{\textwidth}{!}{
			\begin{tabular}{l|cccccccccc}
				\toprule
				& \multicolumn{1}{l}{Caffe  GTX Titan} & \multicolumn{1}{l}{Caffe GTX 980} & \multicolumn{1}{l}{CNTK  GTX Titan} & \multicolumn{1}{l}{CNTK GTX 980} & \multicolumn{1}{l}{MXNet  GTX Titan} & \multicolumn{1}{l}{MXNet GTX 980} & \multicolumn{1}{l}{TF  GTX Titan} & \multicolumn{1}{l}{TF GTX 980} & \multicolumn{1}{l}{Torch  GTX Titan} & \multicolumn{1}{l|}{Torch GTX 980} \\
				\midrule
				FCN 512 & 1.084\% & -1.901\% & 0.077\% & 4.299\% & 3.157\% & 1.047\% & -5.139\% & 2.944\% & 6.891\% & 4.096\% \\
				FCN 1024 & -0.313\% & -1.084\% & -0.923\% & 4.814\% & -0.315\% & 0.115\% & 2.966\% & 3.134\% & 8.641\% & 4.611\% \\
				FCN 2048 & 1.033\% & 0.006\% & -1.993\% & 4.540\% & 0.190\% & 0.324\% & -2.013\% & 4.165\% & 9.615\% & 4.863\% \\
				\midrule
				\midrule
				Alexnet 256 & -0.848\% & -3.308\% & -0.357\% & 1.397\% & -1.603\% & 0.193\% & -0.473\% & 1.166\% & 3.369\% & 2.007\% \\
				Alexnet 512 & -0.241\% & 0.705\% & -5.474\% & 1.619\% & -1.799\% & -0.397\% & 0.028\% & 2.336\% & 3.858\% & 1.978\% \\
				Alexnet 1024 & 0.535\% & -0.061\% & -1.222\% & 1.996\% & -3.607\% & -0.714\% & 0.388\% & 1.182\% & 3.767\% & 1.771\% \\
				\midrule
				\midrule
				Resnet 16 & -0.282\% & -0.552\% & -3.189\% & -0.704\% & -0.043\% & 3.014\% & -0.581\% & 2.001\% & 1.407\% & -0.300\% \\
				Resnet 32 & -0.010\% & -0.467\% & 0.120\% & 0.108\% & 0.291\% & 0.689\% & 3.108\% & 2.143\% & 0.308\% & 0.061\% \\
				Resnet 64 & 0.394\% & -0.043\% & -0.173\% & -0.119\% & 0.940\% & 1.174\% & 2.524\% & 0.905\% & 0.146\% & 0.174\% \\
				\midrule
				\midrule
				LSTM128 & Not Support & Not Support & -3.058\% & 0.089\% & -1.390\% & -0.399\% & 1.084\% & 0.540\% & 0.929\% & 0.080\% \\
				LSTM256 & Not Support & Not Support & 1.326\% & -0.208\% & 3.870\% & 0.024\% & 0.708\% & 0.429\% & 0.998\% & 0.031\% \\
				LSTM512 & Not Support & Not Support & 3.486\% & 0.310\% & -4.484\% & 0.339\% & 0.759\% & 0.558\% & -1.237\% & Not Support \\
				\bottomrule
		\end{tabular}}
	\end{table*}
	
	\subsection{I/O Performance}
	\subsubsection{dd Test}
	We use dd command to sequentially write several files of different sizes to hard disks and collect speed information from outputs.
	\begin{table}[H]
		\centering
		\caption{dd Write Test}
		\label{ddwrite}
		\begin{tabular}{@{}lcc@{}}
			\toprule
			& PC Platform \begin{math}
			Diff_\%
			\end{math} & Server Platform \begin{math}
			Diff_\%
			\end{math}\\ \midrule
			1MiB file   & -4.263\% & -4.023\%    \\
			100MiB file & -1.408\% & 6.893\%     \\
			512MiB file & 0.000\%  & 5.714\%     \\
			1GiB file   & 0.636\%  & 3.448\%     \\ \bottomrule
		\end{tabular}
	\end{table}

	From Table \ref{ddwrite} we can see that I/O speeds are very close when comparing docker containers and host systems. Accessing smaller files tends to have a higher speed in docker containers, whilst accessing large files in the host system are a little bit better than in docker containers. We also measure the standard deviation among each set of 20-runs to estimate the stability as illustrated in Table \ref{ddresulttable}. The values of standard deviation from docker container and host system are very close. We can claim that docker containers are as stable as host system in sequentially writing out large files.

	\subsubsection{ioping Test}
	To measure the random access speed, we make use of \textit{ioping} command. The results are rather interesting compared with \textit{dd} tests. As we can see in Table \ref{iopingresult}, random I/O access results are found to be close in our server platform. However, we observe huge performance differences in our PC platform. Docker containers are extremely faster than host system in direct I/O test and I/O latency test. We go deeper into this phenomenon and find that the speed of direct I/O random access in docker container is almost the same as accessing the disk cache. 	
	\begin{table}[H]
		\centering
		\caption{dd Sequential Access Test}
		\label{ddresulttable}
		\begin{tabular}{@{}llll@{}}
			\toprule
			& \multicolumn{1}{c}{Best} & \multicolumn{1}{c}{Average} & \multicolumn{1}{c}{STD} \\ \midrule
			\multicolumn{1}{l|}{Server with docker} & 322                      & 306.476                 & 19.6069             \\
			\multicolumn{1}{l|}{Server w/o docker}   & 309                      & 292.333                 & 21.6272             \\
			\multicolumn{1}{l|}{PC with docker}    & 103                      & 100.561                 & 0.97408             \\
			\multicolumn{1}{l|}{PC w/o docker}      & 103                      & 102.04                  & 0.64888            \\ \bottomrule
		\end{tabular}
	\end{table}

	The results in Table \ref{iopingresult} show that docker container reacts to I/O request almost 100\% faster than it is in the host system. We take a look at the specification of our hard drive in GPU server and notice that it come equipped with Multi-Tier Caching Technology(MTC). Basically in addition to the actual spinning disk storing data, there are multiple layers of NAND Flash installed on the hard drive for quick access of frequently used data for ongoing processes. Docker containers depend on docker engine running in the background taking advantage of MTC for fast small data access and IO requests. Because of that, ioping doesn't go into the actual disk of hard drive during ioping tests and that's why direct IO random access speed and IO latency time have such huge performance difference.
	\begin{table}[H]
	\centering
	\caption{ioping Test Result}
	\resizebox{\textwidth/2}{!}{
		\begin{tabular}{@{}lcc@{}}
			\toprule
			& Server \begin{math}
			Diff_\%
			\end{math} & PC \begin{math}
			Diff_\%
			\end{math} \\ \midrule
			Cache IO random  & 17.447\%      & 23.065\%   \\ access speed \\
			Direct IO random  & -2.521\%      & -282.075\% \\ access speed\\
			IO latency time             & 3.509\%       & 98.999\%   \\
			\bottomrule
		\end{tabular}
	}
	\label{iopingresult}
	\end{table}
	\subsection{Deep Learning Tools}
	After we evaluate individual factors that may affect the performance of deep learning tools, we then test each deep learning tool by training different types of neural networks. Table \ref{dlresult} shows the results of all networks trained by all the tools. Different tools have their own metrics of measuring performance, so we convert them  into the time they take to train one batch of data. The left most column indicates the type of neural network and the number of samples we put into each batch.

	Note that positive numbers in Table \ref{dlresult} illustrate that the docker container outperforms our host system because the host system needs more time to train one batch of data. As Table \ref{dlresult} shows that deep learning tools perform well in docker container overall. For large networks Resnet and LSTM, we can see that the performance of deep learning tools in docker container is as good as in the host system. In those cases that docker container runs slower than the host system, the difference is usually within 5\%. As for smaller networks like FCN and Alexnet, we also find that the performance of each tool running in docker container and in the host system are similar in terms of computational time costs. An interesting phenomenon is found on MXNet, whose training time of the first epoch in docker container is much shorter than the host, as shown in Table \ref{mxnetcompare}. Notice that starting from the second epoch, the difference drops back to normal. This is because the first epoch includes the initial I/O time which has different performance under docker container and host system, while later on the I/O time is hidden by the computing time. These deep learning frameworks implement parallel data loading that data for the next run are pre-loaded during the training process, so that the I/O time is covered by computing time and we don't see much difference after the first epoch.
	\begin{table}[H]
		\centering
		\caption{MXNet 1st Epoch vs. 2nd Epoch}
		\begin{tabular}{l|cc}
			\toprule
			MXNet GTX980 & \multicolumn{1}{c}{Epoch 1 $Diff_\%$} & \multicolumn{1}{c}{Epoch 2 $Diff_\%$} \\
			\midrule
			\midrule
			FCN 512 & 75.707\% & -1.654\% \\
			FCN 1024 & 77.169\% & 0.000\% \\
			FCN 2048 & 74.563\% & 1.720\% \\
			\midrule
			\midrule
			Alexnet 256 & 69.468\% & -0.798\% \\
			Alexnet 512 & 68.451\% & 0.145\% \\
			Alexnet 1024 & 68.871\% & -0.428\% \\
			\bottomrule
		\end{tabular}%
		\label{mxnetcompare}%
	\end{table}%

	\section{Conclusion and future work}\label{conclusion}
	In conclusion, even though there are extra layers lying between applications and hardware resources by using docker containers, docker engine manages to minimize the overhead pretty well. We don't find noticeable drawbacks of docker containers in CPU and GPU tests. Testing programs running in docker containers perform just as good as in the host system. So putting deep learning tools into docker containers is a feasible solution that we can benefit from its flexibility, lightweight, and resource isolation abilities. Different deep learning tools or the same tools with different version numbers can coexist in one system yet maintaining good performance. System administrators of shared servers and cloud platforms can install docker engine in the system and let users download and run their desired images by their own. System administrators can prepare docker images with deep learning tools pre-installed and properly configured so that users only need to focus on their models and algorithms without getting annoyed by dependencies and environment settings.
	
	In the future, this work can be further extended on multiple machines that train large-scale neural networks on a cluster to gain even more acceleration. In this situation, data transmission efficiency needs to be tested among docker containers located in different physical machines. Even within the same node there can be multiple GPUs installed, efficiently making use of more than one GPU at the same time is also important in deep learning. On the other hand, more types of hardware platform can be included. In our work, we mainly focus on the combination of CPU + NVIDIA GPUs. There are other accelerators such as AMD GPUs and Intel Xeon Phi processors. The results we get from our docker container are based on NVIDIA docker which has official support directly from the GPU manufacturer. Whether it is efficient or not to invoke computing devices from other kind needs to be further studied.
	
	\section{Acknowledgement}
The authors would like to thank all the reviewers for their insightful comments and valuable suggestions. This work is supported by Shenzhen Basic Research Grant SCI-2015-SZTIC-002.
	
	\IEEEtriggercmd{\enlargethispage{-5in}}

	\bibliographystyle{IEEEtran}
	\bibliography{references}
	
\end{document}